\documentclass[10pt,a4paper]{article}

\usepackage[english]{babel}
\usepackage{amsmath,amsthm,amssymb,graphicx}
\usepackage{cite}

\newtheorem{prop}{Proposition}[section]

\newtheorem{definition}[prop]{Definition}

\numberwithin{equation}{section}

\title{The universal Airy$_1$ and Airy$_2$ processes in the Totally Asymmetric Simple Exclusion Process}
\author{Patrik L. Ferrari\\ Technische Universit\"at M\"unchen \\e-mail: ferrari@ma.tum.de}

\date{\today}

\begin{document}
\maketitle \sloppy

\begin{abstract}
In the totally asymmetric simple exclusion process (TASEP) two processes arise in the large time limit: the Airy$_1$ and Airy$_2$ processes. The Airy$_2$ process is an universal limit process occurring also in other models: in a stochastic growth model on $1+1$-dimensions, 2d last passage percolation, equilibrium crystals, and in random matrix diffusion. The Airy$_1$ and Airy$_2$ processes are defined and discussed in the context of the TASEP. We also explain a geometric representation of the TASEP from which the connection to growth models and directed last passage percolation is immediate.
\end{abstract}

\section{Introduction}\label{SectIntro}
In mathematics and physics, there are many situations where an observable behaves, in an appropriate limit, as a Gaussian random variable. The most famous example in mathematics is arguably the central limit theorem for the sum of  i.i.d.\ random variables provided the existence of more than two moments. This is the simplest example of universality class and extends in some cases to dependent random variables too.

The Gaussian is however not the only universality class. In the last decade a lot of progress in understanding universality related to random matrix ensembles have been achieved. In 1994 Tracy and Widom determined the limit distribution of the largest eigenvalue in the Gaussian Unitary Ensemble (GUE) of random matrices~\cite{TW94}, thus called GUE Tracy-Widom distribution, $F_2$. Then, in 1999 Baik, Deift, and Johansson proved that the same distribution asymptotically describes the longest increasing subsequence of a random permutation~\cite{BDJ99}. Johansson figured out that $F_2$ occurs also in the shape fluctuation in a point-to-point directed last passage percolation model~\cite{Jo00}. The same limit distribution arises in a stochastic growth model on a one-dimensional substrate, the polynuclear growth model~\cite{PS00}, in other corner growth models~\cite{GTW00}, directed percolation~\cite{BS04,BM04,Jo00}, in vicious random walks~\cite{NKT02}, in non-colliding Brownian motions~\cite{KNT03}, and in the totally asymmetric simple exclusion process (TASEP)~\cite{Jo00,BR00,NS04}. Concerning random matrices, it was meanwhile proven that $F_2$ occurs beyond the GUE ensemble~\cite{Sos99,DKMVZ99}.

In several of the models analyzed, a natural question in an extension to joint distributions. In random matrices, the extension is Dyson's Brownian Motion~\cite{Dys62}. The process describing the evolution of the largest eigenvalue in $\beta=2$ Dyson's Brownian Motion is, in the limit of large matrix dimension, the Airy process (below referred as Airy$_2$ process). This process was discovered by Pr\"ahofer and Spohn in a growth model of a surface, the polynuclear growth (PNG) model~\cite{PS02}, where the Airy$_2$ process describes the interface for large growth time $t$. As expected, it arises also in discretized versions of the model~\cite{Jo03b,SI03,SI04}. The PNG surface can be also related to an evolution of certain Young tableaux, in terms of which the result has been extended by Borodin and Olshanski~\cite{BO04}. Discrete versions of the PNG model are tightly linked with a point-to-point directed polymer problem (or directed last passage percolation) where again the Airy$_2$ process is a limit process~\cite{Jo03b}. A different class of models where the Airy$_2$ process arises are tiling models, the most studied one is the Aztec diamond~\cite{Jo03}, which can be mapped to the six-vertex model at the free-fermion point~\cite{FS06}. Finally, the Airy$_2$ process occurs in equilibrium statistical mechanics too. It describes the fluctuation of the border of a facet in the 3d-Ising corner model~\cite{FS03,OR01}, but the same result is expected to hold in much more generality for short-range interactions crystals at low temperature~\cite{FPS03}, see also~\cite{KOS03} for related studies on the macroscopic shapes.

We have seen that the GUE Tracy-Widom distribution and the Airy$_2$ process appears in several models. How stable are they under change of initial conditions or intrinsic symmetries of the model? While the scaling exponents remain universal, the precise distribution depends on initial conditions or symmetries of the model. However, the results obtained so far indicate that the limit distribution is still universal but one has to divide the universality class in a few subclasses.

In classical random matrix theory, besides the GUE, there are the Gaussian Orthogonal Ensemble (GOE) and the Gaussian Symplectic Ensemble (GSE)~\cite{Dys62b}. They were introduced as an approximation to the real Hamiltonian of heavy nuclei and the different ensembles reflect the different intrinsic symmetries of the system under consideration (e.g., the time-reversal symmetry, which is broken by external magnetic fields). With the same scaling as for GUE, Tracy and Widom obtained the limit distribution of the largest eigenvalues for GOE and GSE too~\cite{TW96,TW04}. This is a clear example of the stronger stability of the scaling exponent (in this case the fluctuation exponent) v.s.\ the distribution functions. In the PNG model, the GOE Tracy-Widom distribution, $F_1$, occurs when the growth is on a flat substrate~\cite{PS00,BR99b}. The connection between the PNG model and the GOE ensemble actually goes beyond the largest eigenvalue~\cite{Fer04}.

Since 2002, the question of the analogue of the Airy$_2$ process in this setting was open and, for some aspects, like in random matrix, is still open. In 2005 Sasamoto discovered a new process in the context of the TASEP~\cite{Sas05}, see~\cite{BFPS06} for a complete derivation. This process is now called the Airy$_1$ process and, although has a similar mathematical structure as the Airy$_2$ process, a lot of the probabilistic interpretation that was present for the latter is lost (or, maybe, not yet understood). The first generalization of the setting considered in~\cite{Sas05} has been carried out in~\cite{BFP06}. The evidence of universality of the Airy$_1$ process is not yet as large as for the Airy$_2$ process, but it is expected to occur in point-to-line directed last passage percolation, in the polynuclear growth model on a flat (not necessarily horizontal) substrate, and possibly also in GOE Dyson's Brownian Motion.

In this paper we first define the Airy$_1$ and Airy$_2$ processes and give their known properties. Then we introduce the continuous time TASEP and explain under which initial conditions the two Airy processes occurs. For the interested reader around some of the methods used in relation to the Airy$_2$ process as well as related models, we refer to a the lecture notes~\cite{Spo05,Jo05}, and the surveys~\cite{FP05,De06,Lyo03,BKPV05,Sos06}.

\section{The Airy$_1$ and Airy$_2$ processes}\label{SectProcesses}
In this section we define the two Airy processes and give the known properties.
\begin{definition}[The Airy$_1$ process]
The \emph{{Airy$_1$ process}} ${\mathcal{A}}_{\rm 1}$ is the process with $m$-point joint distributions at
\mbox{$u_1< u_2< \ldots < u_m$} given by the Fredholm determinant
\begin{equation}\label{eqF1}
\mathbb{P}\Big(\bigcap_{k=1}^m\{{\mathcal{A}}_{\rm 1}(u_k)\leq s_k\}\Big)=
\det(\mathbf{1}-\chi_s K_1\chi_s)_{L^2(\{u_1,\ldots,u_m\}\times\mathbb{R})}
\end{equation}
where $\chi_s(u_k,x)=\mathbf{1}(x>s_k)$ and the kernel $K_1$ is a follows. Let \mbox{$A_1(x,y)=\mathrm{Ai}(x+y)$} and $H_1=-\Delta$ with $\Delta$ the one-dimensional Laplacian. Then the $K_1$ is defined by
\begin{equation}\label{eqKernelF1}
K_1(u,s;u',s')=-(e^{-(u'-u)H_1})(s,s')\mathbf{1}(u<u')+(e^{u H_1} A_1 e^{-u' H_1})(s,s').
\end{equation}
\end{definition}

\begin{definition}[The Airy$_2$ process]
The \emph{{Airy$_2$ process}} ${\mathcal{A}}_{\rm 2}$ is the process with $m$-point joint distributions at
\mbox{$u_1< u_2< \ldots < u_m$} given by the Fredholm determinant
\begin{equation}\label{eqF2}
\mathbb{P}\Big(\bigcap_{k=1}^m\{{\mathcal{A}}_{\rm 2}(u_k)\leq s_k\}\Big)=
\det(\mathbf{1}-\chi_s K_2\chi_s)_{L^2(\{u_1,\ldots,u_m\}\times\mathbb{R})}
\end{equation}
where $\chi_s(u_k,x)=\mathbf{1}(x>s_k)$ and $K_2$ is the extended Airy kernel.
Let $H_2$ be the Airy operator
\begin{equation}
H_2 = -\frac{d^2}{dx^2}+x
\end{equation}
and $A_2$ the (one-time) Airy kernel with entries
\begin{equation}
A_2(x,y)=\int_{\mathbb{R}_+}\mathrm{d}\lambda \mathrm{Ai}(x+\lambda)\mathrm{Ai}(y+\lambda).
\end{equation}
Then the define the kernel
\begin{equation}\label{eqKernelF2}
K_2(u,s;u',s')=-(e^{-(u'-u)H_2})(s,s')\mathbf{1}(u<u')+(e^{u H_2} A_2 e^{-u' H_2})(s,s').
\end{equation}
\end{definition}

\subsubsection*{Explicit expressions}
The explicit formulas of the $K_1$ and $K_2$ kernels are the following. For the Airy$_1$ process, as shown in Appendix A of~\cite{BFPS06}, one has
\begin{eqnarray}\label{eqKernelF1Expanded}
& &\hspace{-2em} K_1(u,s;u',s')=-\frac{1}{\sqrt{4\pi (u'-u)}}\exp\left(-\frac{(s'-s)^2}{4 (u'-u)}\right) \mathbf{1}(u<u') \nonumber \\
& &\hspace{-2em} + \mathrm{Ai}(s+s'+(u'-u)^2) \exp\left((u'-u)(s+s')+\frac23(u'-u)^3\right),
\end{eqnarray}
with $\mathrm{Ai}$ the Airy function~\cite{AS84}, while the kernel for the Airy$_2$ process writes~\cite{PS02,Jo03b}
\begin{equation}\label{eqKernelF2Expanded}
K_2(u,s;u',s')=\left\{\begin{array}{ll}
\int_{\mathbb{R}_+}\mathrm{d}\lambda e^{(u'-u)\lambda}\mathrm{Ai}(x+\lambda)\mathrm{Ai}(y+\lambda),&u\geq u',\\[0.5em]
-\int_{\mathbb{R}_-}\mathrm{d}\lambda e^{(u'-u)\lambda}\mathrm{Ai}(x+\lambda)\mathrm{Ai}(y+\lambda),&u<u'.
\end{array}\right.
\end{equation}

\subsubsection*{One-point distributions}
The Airy$_1$ and Airy$_2$ processes are stationary and their one-point distribution functions are given in tems of the GOE and GUE Tracy-Widom distributions of random matrices~\cite{TW96,TW94}. Namely, as shown in~\cite{FS05b},
\begin{equation}
\mathbb{P}({\mathcal{A}}_{\rm 1}(0)\leq s)=F_1(2s),
\end{equation}
and, for the Airy$_2$ process~\cite{PS02},
\begin{equation}
\mathbb{P}({\mathcal{A}}_{\rm 2}(0)\leq s)=F_2(s).
\end{equation}

\subsubsection*{Spatial correlations}
Some information on the spatial correlation have been already determined for the Airy$_2$ process. Locally it behaves like a diffusion and on a long distance the spatial correlations have a slow polynomial decay. More precisely, define the function $g$ by
\begin{equation}
\mathrm{Var}({\mathcal{A}}_{\rm 2}(u)-{\mathcal{A}}_{\rm 2}(0)) = g(u).
\end{equation}
$g$ grows linearly for small $u$ and that the Airy process has long range correlations~\cite{PS02}:
\begin{equation}
g(u)= \left\{\begin{array}{ll}
2u+{\mathcal{O}}(u^2)&\textrm{for }|u|\textrm{ small,}\\
g(\infty)-2u^{-2}+{\mathcal{O}}(u^{-4})&\textrm{for }|u|\textrm{ large.}
\end{array}\right.
\end{equation}
with $g(\infty)=1.6264\ldots$. The coefficient $2$ of the correlation's decay is determined in~\cite{AvM03,Wid03}.
For the Airy$_2$ process a set of PDE's~\cite{AvM03} and ODE's~\cite{TW03} have been obtained.

For the Airy$_1$ process such analysis is still missing, but it is expected that the global behavior of $g$ is the same, as indicated also from some preliminaly numerical computations~\cite{Sas05}.

\subsubsection*{On the Fredholm determinants}
The joint distributions for the Airy processes are given in terms of Fredholm determinants. Are they really well defined? One way of considering the Fredholm determinants (\ref{eqF1}) and (\ref{eqF2}) is simply via their Fredholm series expansion, namely
\begin{eqnarray}\label{eqFredholm}
& &\det(\mathbf{1}-\chi_s K\chi_s)_{L^2(\{u_1,\ldots,u_m\}\times\mathbb{R})} \\
&=& \sum_{n\geq 0}\frac{(-1)^n}{n!}\sum_{i_1,\ldots,i_n=1}^{m} \int_{x_1\geq s_{i_1}} \hspace{-1.5em}\mathrm{d}x_1 \cdots \int_{x_n\geq s_{i_n}} \hspace{-1.5em} \mathrm{d}x_n \det(K(u_{i_k},x_k;u_{i_l},x_l))_{1\leq k,l \leq n}. \nonumber
\end{eqnarray}
One can see that (\ref{eqFredholm}) is absolutely summable/integrable for the kernel $K=K_1$ and $K=K_2$. However, it is sometimes useful to consider the expressions (\ref{eqF1}) and (\ref{eqF2}) as Fredholm determinant of an operator on the Hilbert space \mbox{$\mathcal{H}=L^2(\{u_1,\ldots,u_m\}\times\mathbb{R})$}. The Fredholm determinant is well defined for trace class operators.

For (\ref{eqF2}) there is no problem, since $K_2$ is trace class on $\mathcal{H}$~\cite{TW94}. This is not the case for $K_1$. In fact, due to contribution of the diffusion part of the kernel, $K_1$ is not even Hilbert-Schmidt on $\mathcal{H}$. However, as shown in Appendix A of~\cite{BFP06}, there exists a conjugate operator of $K_1$ that is trace class on $\mathcal{H}$. With conjugate operator we mean an operator $\widetilde K_1$ that leads to the same correlation functions and Fredholm series expansion, i.e., such that
\begin{equation}
\det(\widetilde K_1(u_{i_k},x_k;u_{i_l},x_l))_{1\leq k,l \leq n}=\det(K_1(u_{i_k},x_k;u_{i_l},x_l))_{1\leq k,l \leq n}
\end{equation}
holds. Equivalently, one can employ a weighted Hilbert space where $K_1$ is trace class.

\section{The TASEP}\label{SectTASEP}
The totally asymmetric simple exclusion process (TASEP) is an interacting stochastic particle system. It consists on particles on $\mathbb{Z}$ with the exclusion constraint that at any given time $t$, every site $j\in\mathbb{Z}$ can be occupied at most by one particle. Thus a configuration of the TASEP can be described by \mbox{$\eta=\{\eta_j,j\in\mathbb{Z}|\eta_j\in\{0,1\}\}$}. $\eta_j$ is called the \emph{occupation variable} of site $j$, which is defined by $\eta_j=1$ if site $j$ is occupied and $\eta_j=0$ if site $j$ is empty.

\begin{figure}
\begin{center}
\includegraphics{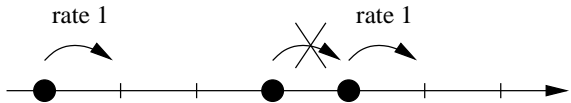}
\caption{Dynamics of the TASEP. Particles jump with rate one on the right site, but under the constraint that the site is empty.}
\label{FigDynamics}
\end{center}
\end{figure}

The dynamics of the TASEP is defined as follows. Particles jumps on the neighboring right site with rate $1$ provided that the site is empty, see Figure~\ref{FigDynamics}.
This means that jumps are independent of each other and are performed after an exponential waiting time with mean $1$, which starts from the time instant when the right neighbor site is empty. More precisely, let $f$: $\Omega\to \mathbb{R}$ be a function depending only on a finite number of $\eta_j$'s. Then the backward generator of the TASEP is given by
\begin{equation}\label{1.1}
Lf(\eta)=\sum_{j\in\mathbb{Z}}\eta_j(1-\eta_{j+1})\big(f(\eta^{j,j+1})-f(\eta)\big).
\end{equation}
Here $\eta^{j,j+1}$ denotes the configuration $\eta$ with the occupations at sites $j$ and \mbox{$j+1$} interchanged. The semigroup $e^{Lt}$ is well-defined as acting on bounded and continuous functions on $\Omega$. $e^{Lt}$ is the transition probability of the TASEP~\cite{Li99}.

\subsubsection*{Observables}
Since the dynamics does not interchange particles, they can be labelled by an index $k\in I$, $I\subset \mathbb{Z}$. We denote by $x_k(t)$ the positions of the particle $k$ at time $t$ and use the right-left ordering, i.e., $x_{k+1}(t)\leq x_k(t)$, for all $t$. There are two observables (quantity of interest) which are closely related: the position of given particles and the integrated current at fixed locations.

The TASEP integrated current at position $x$ and time $t$, $J(x,t)$, is the number of particles which jumped from site $x$ to site $x+1$ during the time interval $[0,t]$. Let us label by $1$ the right-most particle starting at position $x_1(0)\leq x$. Then $J(x,t)$ and $x_s(t)$ are related by
\begin{equation}
\mathbb{P}(J(x,t)\geq s)=\mathbb{P}(x_s(t)\geq x+1).
\end{equation}

Below we consider as observable a given subset of particles. For some finite $I\subset \mathbb{Z}$, we explain in which scaling limit the joint distributions of
\begin{equation}
\{x_k(t),k\in I\}
\end{equation}
are governed by the Airy processes.

\subsubsection*{Step initial conditions and Airy$_2$ process}
The Airy$_2$ process appears, is called step-initial condition. The initial configuration is deterministic defined by
\begin{equation}
x_k(t=0)=-k,\quad k\in \mathbb{Z}_+.
\end{equation}
First consider the macroscopic behavior, around which one analyzes the fluctuations. The macroscopic limit density $u(\xi)$ is given by
\begin{equation}
u(\xi)=\frac{\mathrm{d}}{\mathrm{d}\xi}\lim_{t\to\infty} t^{-1}\mathbb{E}\big(\#(k: x_k(t)\geq \xi t)\big).
\end{equation}
It was proven by Rost~\cite{R81} that the macroscopic density has a linearly decreasing region, namely
\begin{equation}
u(\xi)=\left\{
\begin{array}{ll}
1,&\xi\leq -1,\\
1-(\xi+1)/2,&-1\leq \xi\leq 1,\\
0,&\xi\geq 1.
\end{array}
\right.
\end{equation}

The particles which build the bulk of the linear region are the one with particles number $\alpha t$, $\alpha \in (0,1)$. First it was shown, by means of a growth model~\cite{Jo00b}, that the fluctuations of their position is $F_2$-distributed. Then Johansson in~\cite{Jo03b} proves a functional limit theorem in a discrete-time setting.
Its continuous-time version writes
\begin{equation}
\lim_{t\to\infty}\frac{x_{[t/4+u(t/2)^{2/3}]}(t)+2u(t/2)^{2/3}-u^2(t/2)^{1/3}}{-(t/2)^{1/3}}={\mathcal{A}}_{\rm 2}(u).
\end{equation}

\subsubsection*{Alternating initial conditions and Airy$_1$ process}
The situation in which the Airy$_1$ process was first discovered is the alternating initial condition. It is the deterministic initial configuration given by
\begin{equation}
x_k(t=0)=-2 k,\quad k\in \mathbb{Z}.
\end{equation}

The macroscopic density is simply $u(\xi)=1/2$ and particles moves with average speed $1/4$. For alternating initial condition, it was expected by universality that the one-point distribution was $F_1$-distributed. It was the case for a continuous version of a corresponding growth model~\cite{BR00,PS00}.

The extension to joint distribution remained for a while unsolved. The solution came from a new approach. Starting from a determinantal formula by Sch\"utz~\cite{Sch97} of the joint distribution of particle positions, Sasamoto~\cite{Sas05} found a clever way of rewriting in term of a \emph{signed} determinantal point process on a larger set of variables. This is explained in details in~\cite{BFPS06}, where it is also proven the pointwise convergence of kernel to $K_1$. The finite-dimensional convergence to the Airy$_1$ process has been obtained for a discrete-time version of the TASEP in~\cite{BFP06}. Its continuous-time analogue writes
\begin{equation}
\lim_{t\to\infty}\frac{x_{[t/4+ut^{2/3}]}(t)+2ut^{2/3}}{-t^{1/3}}={\mathcal{A}}_{\rm 1}(u).
\end{equation}

\subsubsection*{Stationary TASEP}
Besides deterministic initial conditions, one can also consider the stationary TASEP. The translation invariant measures for the stationary TASEP are the Bernoulli-$\rho$ measures, $\rho\in[0,1]$, i.e., $\mathbb{P}(\eta_k=1)=\rho$ and the random variables $\eta_k$'s are independent. Between other quantities, an explicit expression for the one-point distribution has been obtained in continuous time and discrete time TASEP with parallel update~\cite{BR00,PS01,FS05a}.

\subsubsection*{Universality and generalizations}
By universality it is expected that the Airy$_1$ process is the large time limit process for a larger class of models. First of all, the Airy$_1$ process should appear for all deterministic and periodic initial conditions for the TASEP, and its discrete time versions. It is also expected that the Airy processes occur in the partially asymmetric simple exclusion process (PASEP), as soon as the drift is non-zero. As it will be explained below, the TASEP can be reinterpreted as a growth model belonging to the KPZ universality class. For model in this class of universality, we expect that the Airy processes appears too. More precisely, the Airy$_1$ process for growth on a flat substrate and the Airy$_2$ process for growth leading a curved shape due to initial conditions. Some of these growth models already studied are directly linked with directed last passage percolation. Thus, we expect the Airy$_1$, resp.\ Airy$_2$, process to describe the process associated to point-to-line, resp.\ point-to-point, directed last passage percolation.

As discussed briefly in the Introduction, the Airy$_2$ process has been obtained in several models. For the Airy$_1$ process only in a few cases the analysis is available so far. The first generalizations has been made in~\cite{BFP06}, where the TASEP is analyzed for a larger set of initial conditions and in the discrete-time TASEP with sequential update. This update rule is as follows. At each time step particles are processed sequentially from right to left, i.e., starting from right to left, if the site on the right of a particle is empty, then this particle jumps there with probability $p$. This update rule allows to shift blocks of particles to the right during one time-step.

More precisely, in~\cite{BFP06} it is considered the deterministic and periodic initial configuration where particles start from $d\mathbb{Z}$ for any $d\geq 2$ fixed. Then, it is proven that the particle positions converges, properly rescaled, to the Airy$_1$ process. More precisely,
\begin{equation}
\lim_{t\to\infty}\frac{x_{n(u,t)}(t)-\mu u t^{2/3}}{-\kappa t^{1/3}}={\mathcal{A}}_{\rm 1}(u),
\end{equation}
where the constants $\mu$ and $\kappa$ are given by
\begin{equation}
\kappa=\frac{(2(1-p)p)^{1/3}(d(d-1))^{2/3}}{d-p},\quad \mu=-\kappa^2 \frac{2}{d-1}
\end{equation}
and the index of the particle $n(u,t)$ by
\begin{equation}
n(u,t)=\bigg[\frac{p(d-1)}{d(d-p)}t-\frac{\mu u}{d}t^{2/3}\bigg].
\end{equation}
The convergence is in the sense of finite-dimensional distributions.

In discrete time, a second natural update rule is the parallel update. It consists in first checking for all particles if they can jump (i.e., if their right-neighbor is empty) and then, simultaneously and independently, these particles jump to the right with probability $p$. Some progress in the parallel update has been recently made in~\cite{PP06}, but there are not yet results concerning the limit processes. Other update rules can also be considered, see the review~\cite{Sch00}.

A natural extension is to consider the partially asymmetric simple exclusion process (PASEP), but very few is known. The scaling exponent have been determined also on a rigorous level~\cite{BS06}, but most of the methods applied for the TASEP do not easily extend to the PASEP. Also in this model the Airy$_1$ and Airy$_2$ processes are expected to occur as soon as the drift is non-zero.

\subsubsection*{A geometric representation}
In this final part we present a geometric representation of the TASEP, from which the reason why the Airy$_1$ process should occur in the point-to-line last passage directed percolation is apparent.

\begin{figure}[h!]
\begin{center}
\includegraphics{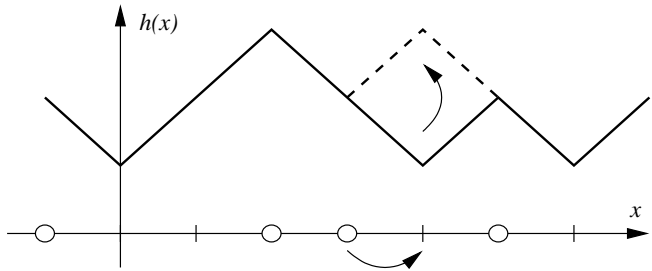}
\caption{The interface (height line) associated with a TASEP configuration.}
\label{FigGrowthModel}
\end{center}
\end{figure}

\vspace{6pt}\noindent\emph{Growth model.} Given a configuration of particles, one associates a height function $h$. Set the height at a given position, e.g., $h(0)=0$. Then, the height function at $x\in\mathbb{Z}$ is obtained as follows. The height differences are given by the occupation variables of the TASEP, more precisely, $h(x+1)-h(x)=1-2\eta_x$. Thus, starting from $x=0$, one can define the height $h(x)$, $x\in\mathbb{Z}$. The extension to $x\in\mathbb{R}$ is just by linear interpolation of the heights on $\mathbb{Z}$. An example is shown in Figure~\ref{FigGrowthModel}.

The dynamics of the TASEP is reflected in the height function picture by a stochastic growth in the vertical direction. In fact, when a particle jumps from $x$ to $x+1$, in the height function a valley $\diagdown\hspace{-0.1em}\diagup$ becomes a mountain $\diagup\hspace{-0.1em}\diagdown$ and the height at $x+1$ is increases by two, see Figure~\ref{FigGrowthModel}. In this way, the TASEP can be equivalently seen as a stochastic growth model of an interface.

\begin{figure}[h!]
\begin{center}
\includegraphics{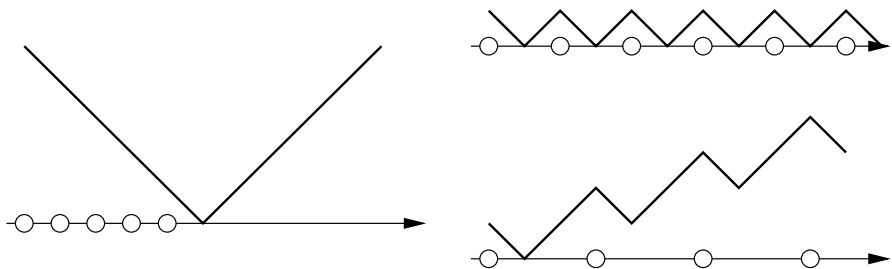}
\caption{Interfaces associated with a step initial configuration (left) and two periodic initial configurations (right).}
\label{FigInitialConfigurations}
\end{center}
\end{figure}

The two types of initial conditions for the TASEP discussed above corresponds to the so-called corner and flat growth. More precisely, step initial conditions become growth starting from a corner-like initial surface, while periodic initial conditions become growth from a flat (also tilted) initial configuration, see Figure~\ref{FigInitialConfigurations}. With flat is it meant that the surface is flat after a coarse graining of order one.

Corner growth leads then to a macroscopic shape which is curved and the large time limit process describing the surface is the Airy$_2$ process, while growth on a flat substrate keeps a macroscopically flat interface with fluctuations described by the Airy$_1$ process.

For discrete TASEP with parallel update, by setting space and time units to $p$ and letting $p\to 0$, one obtains the (continuous time) polynuclear growth (PNG) model~\cite{PS00,PS02}, a stochastic growth model in the KPZ class.

\vspace{6pt}\noindent \emph{Directed last passage percolation.}
The TASEP can be mapped to last passage percolation on $\mathbb{Z}^2$ with
i.i.d.\ exponentially distributed random variables $\omega(m,n)$,
$m,n\in\mathbb{Z}$. The precise connection is that $\omega(m,n)$ is the
waiting time of particle number $n$ to jump from position $m-n-1$ to $m-n$.

To explain it better, consider the step-initial condition, which, as we will see, corresponds to point-to-point directed last passage percolation. The last passage time $G(m,n)$ is given by
\begin{equation}\label{eqDP}
G(m,n)=\max_{\pi:(1,1)\to(m,n)}\ell(\pi),\quad \ell(\pi)=\sum_{(i,j)\in\pi}\omega(i,j)
\end{equation}
where $\pi:(1,1)\to(m,n)$ are up-right paths (concatenation of $(0,1)$ or $(1,0)$ steps), going from $(1,1)$ to $(m,n)$. If there are no up-right paths from $(1,1)$ to $(m,n)$, we set $G(m,n)=0$. Then define the domain
\begin{equation}
\mathcal{B}_t=\{(m,n)\in\mathbb{Z}^2| G(m,n)\leq t\}.
\end{equation}
The border of this domain is directly connected with the TASEP as follows.

\begin{figure}[h!]
\begin{center}
\includegraphics{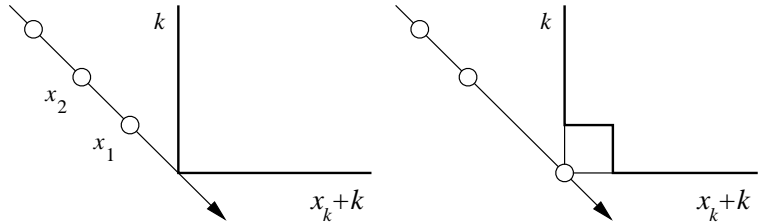}
\caption{TASEP with step-initial condition and directed percolation. Left is the $t=0$ configuration and right is the one at time $t=\omega(1,1)$.}
\label{FigDPtasep}
\end{center}
\end{figure}

To the step-initial configuration $x_k(0)=-k$, $k=1,2,\ldots$, we associate the height line of Figure~\ref{FigInitialConfigurations} and after a clockwise rotation of 45 degrees, one obtain the line of Figure~\ref{FigDPtasep} (left). This line is the border of $\mathcal{B}_0$, denoted by $\partial \mathcal{B}_0$,
which passes through the coordinates $(x_k(0)+k,k)$, $k=1,2,\ldots$. Then, at time $t=\omega(1,1)$, $\partial \mathcal{B}_t$ becomes the one of Figure~\ref{FigDPtasep} (right). This happens exactly when the particle number $1$ jumps from position $-1$ to position $0$. The correspondence actually extends for all $t\geq 0$ and using the identity $G(m,n)=\omega(m,n)+\max\{G(m-1,n),G(m,n-1)\}$ one can see that $G(m,n)$ is the time needed for particle number $n$ to reach site $m-n$.

Finally, periodic initial conditions become the point-to-line problem in last passage percolation. More precisely, the $\pi$ in (\ref{eqDP}) are all paths ending at $(m,n)$ and starting from a given line, e.g., the line is $\{(u,-u),u\in\mathbb{Z}\}$ for $x_k(0)=-2k$, $k\in\mathbb{Z}$.

%\newcommand{\bibliodir}[1]{../../Biblio/#1}
%\bibliographystyle{\bibliodir{patplain}}
%\bibliography{\bibliodir{Biblio}}

\end{document}